\documentclass[conference]{IEEEtran}
\IEEEoverridecommandlockouts
\usepackage{cite}
\usepackage{amsmath,amssymb,amsfonts}
\usepackage{algorithmic}
\usepackage{graphicx}
\usepackage{textcomp}
\usepackage{xcolor}
\usepackage{cite}
\usepackage{amsmath,amssymb,amsfonts}
\usepackage{algorithmic}
\usepackage{graphicx}
\usepackage{textcomp}
\usepackage{times}
\usepackage{soul}
\usepackage{url}
\usepackage{multirow}
\usepackage[utf8]{inputenc}
\usepackage{amsthm}
\usepackage{booktabs}
\usepackage{algorithm}
\usepackage[switch]{lineno}
\usepackage{placeins}
\usepackage{array}
\usepackage{graphicx}
\usepackage{subcaption}
\def\BibTeX{{\rm B\kern-.05em{\sc i\kern-.025em b}\kern-.08em
    T\kern-.1667em\lower.7ex\hbox{E}\kern-.125emX}}
\begin{document}

\title{Exploiting Longitudinal Speech Sessions via Voice Assistant Systems for Early Detection of Cognitive Decline}

\author{
    \IEEEauthorblockN{Kristin Qi\textsuperscript{1}, Jiatong Shi\textsuperscript{2}, Caroline Summerour\textsuperscript{3}, John A. Batsis\textsuperscript{3}, Xiaohui Liang\textsuperscript{1}}
    
    \IEEEauthorblockA{\textsuperscript{1}Computer Science, University of Massachusetts, Boston, MA, USA}
    \IEEEauthorblockA{\textsuperscript{2}Computer Science, Carnegie Mellon University, Pittsburgh, PA, USA}
    \IEEEauthorblockA{\textsuperscript{3}School of Medicine, University of North Carolina, Chapel Hill, NC, USA}
    
    \IEEEauthorblockA{Email: \{yanankristin.qi001, xiaohui.liang\}@umb.edu, \\jiatongs@cs.cmu.edu, caroline\_summerour@med.unc.edu, john.batsis@unc.edu}
}

\maketitle

\begin{abstract} 
Mild Cognitive Impairment (MCI) is an early stage of Alzheimer's disease (AD), a form of neurodegenerative disorder. Early identification of MCI is crucial for delaying its progression through timely interventions. Existing research has demonstrated the feasibility of detecting MCI using speech collected from clinical interviews or digital devices. However, these approaches typically analyze data collected at limited time points, limiting their ability to identify cognitive changes over time. This paper presents a longitudinal study using voice assistant systems (VAS) to remotely collect seven-session speech data at three-month intervals across 18 months. We propose two methods to improve MCI detection and the prediction of cognitive changes. The first method incorporates historical data, while the second predicts cognitive changes at two time points. Our results indicate improvements when incorporating historical data: the average F1-score for MCI detection improves from 58.6\% to 71.2\% (by 12.6\%) in the case of acoustic features and from 62.1\% to 75.1\% (by 13.0\%) in the case of linguistic features. Additionally, the prediction of cognitive changes achieves an F1-score of 73.7\% in the case of acoustic features. These results confirm the potential of VAS-based speech sessions for early detection of cognitive decline.

\end{abstract}

\section{Introduction}
\label{sec: intro}

Alzheimer's disease (AD), a leading cause of death, is a complex neurodegenerative disorder that severely impacts the quality of life. It is characterized by memory loss, cognitive impairment, and behavior changes~\cite{nichols2019global}. While there is no cure for AD, its precursor, Mild Cognitive Impairment (MCI), represents a key stage for early detection and intervention, as interventions at this stage could potentially slow or even halt the progression of AD~\cite{langa2014diagnosis}. Traditional assessments of MCI rely on in-clinic cognitive tests such as the Mini-Mental State Examination (MMSE)~\cite{folstein1975mini}, Montreal Cognitive Assessment (MoCA)~\cite{nasreddine2005montreal}, and the Mini-Cog~\cite{borson2003mini}. Despite their efficacy, their reliance on specialized clinicians for administration limits their feasibility for long-term monitoring.  

Recent studies in speech analysis have demonstrated its potential as a non-invasive and cost-effective method for early detection and monitoring of MCI~\cite{martinez2021ten}. Typically, speech analysis involves identifying signs such as long hesitation gaps, worse articulation, irregular fluency, and unusual word choices~\cite{meteyard2009relation,yuan2021pauses} to differentiate between individuals with MCI and healthy controls (HC). Given that spontaneous speech characteristics contain clinically relevant information for MCI identification~\cite{reilly2010cognition,mueller2018connected}, efforts were made to collect datasets for dementia research, including ADReSS~\cite{luz2020alzheimer}, the Pitt Corpus~\cite{becker1994natural}, and Dem@Care~\cite{karakostas2016care}. Newer methods use smartphones, tablets, and computers for data collection~\cite{maguire2021shaping}. However, these approaches generally compare speech patterns across patients at specific time points rather than observing changes in an individual's speech performance over time~\cite{luz2020alzheimer}. 

The increasing adoption of voice assistant systems (VAS) impacts the lives of the elderly by simplifying daily tasks through speech interaction and enabling hands-free operations~\cite{liang2022evaluating, berridge2021domain}. We present using VAS, such as Amazon Alexa, to collect longitudinal speech session data from the elderly at three-month intervals over 18 months, resulting in seven sessions. This approach aims to monitor speech changes for the early detection of cognitive decline. Specifically, our study engaged participants over the age of 65 in virtual sessions, where they answered 18 cognitive-task questions selected from existing telephone-based cognitive interviews~\cite{brandt1988telephone}. After completing these tasks, participants' cognitive abilities were assessed by the Montreal Cognitive Assessment (MoCA)~\cite{nasreddine2005montreal} scores as the ground-truth diagnosis labels for machine learning model analysis.

We recorded and analyzed the speech data of the 35 participants, both with MCI and HC groups. Throughout the study, we obtained data from 243 sessions. By employing the advantage of longitudinal data, we propose two methods: one to detect participants' cognitive states by incorporating their historical data and another to identify cognitive changes of an individual participant between two time points. We further examine both acoustic and linguistic features extracted from self-supervised learning and large language models, and evaluate results derived from machine learning models.  

Our research contributions are three-fold: i) collecting a longitudinal VAS-based speech session data for early cognitive decline detection and analysis; ii) proposing and evaluating a \textbf{cognitive state detection} method that uses historical data to detect cognitive states (MCI vs. HC); and iii) proposing and evaluating a \textbf{cognitive change prediction} method that predicts changes into three classes (improved, no change, decline) using random pairs of data samples collected at two different time points. The evaluation results confirm the potential utility of VAS-based speech sessions for the early detection of cognitive decline.

\section{Related Work}
\label{sec: related-work}
Recent works show that linguistic and acoustic features are characteristics in automatically detecting cognitive decline~\cite{fraser2016linguistic,mirheidari2021identifying}. These studies demonstrate that training acoustic features from audios and linguistic features from transcripts with machine learning can differentiate between patients with neurodegenerative diseases and healthy individuals. Examples of acoustic features are Mel-Frequency Cepstral Coefficients (MFCC), ComParE~\cite{eyben2013recent}, MRCG~\cite{chen2014feature}, and eGeMAPS~\cite{eyben2015geneva}. Examples of linguistic features are hidden states from pre-trained neural networks, such as BERT~\cite{syed2020automated}, RoBERTa~\cite{ilias2022explainable}, and GloVe~\cite{mirheidari2021identifying}. Machine learning algorithms range from conventional classifiers like Linear Discriminant Analysis (LDA), Decision Trees (DT), and Support Vector Machine (SVM) to advanced neural networks like Convolutional Neural Networks (CNN), Long-Short Term Memory units (LSTM), and Transformer networks. Previous studies have primarily focused on analyzing datasets collected at a single time point. In contrast, we collect data over time using VAS and provide longitudinal data analysis. Additionally, we extend our experiments by incorporating newer models, including pre-trained self-supervised learning models and LLaMA-2.

\begin{figure*}[t]

  \centering

  \includegraphics[width=0.9\textwidth]{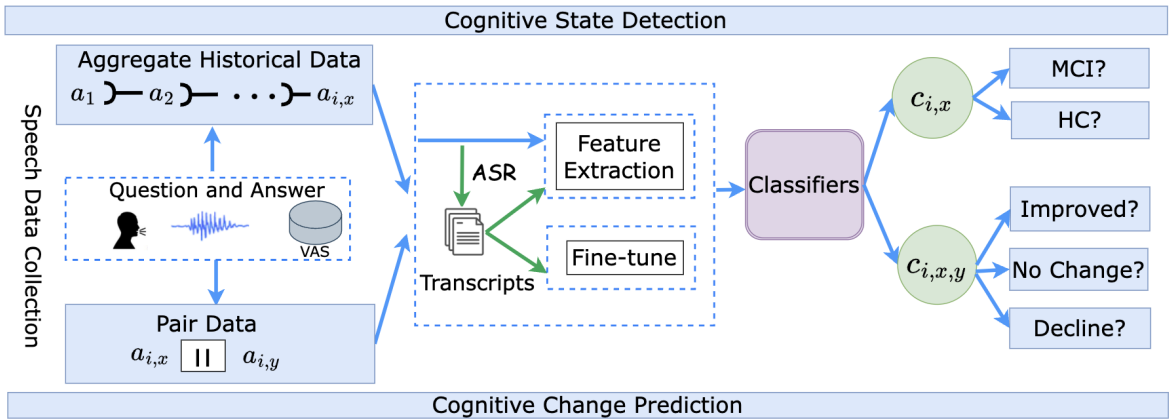}

  \caption{Workflow of the two proposed methods. \textbf{Cognitive state detection} method (top) classifies between MCI and HC by aggregating historical data to predict a participant's current cognitive state. \textbf{Cognitive change prediction} method (bottom) identifies cognitive changes by using randomly paired data from two different sessions.}
  
  \label{fig:pipe}

\end{figure*}

\section{VAS-based speech sessions}
Our study involved 35 participants, consisting of 16 females and 19 males. Among these, 20 were identified with MCI, and 15 were healthy controls (HC). The average age of the participants was 72 (standard deviation $= 5.1$). Table~\ref{table:demo} presents the demographic information exclusively for the first session. Speech was collected through virtual sessions repeated every three months from 2022 to 2024, culminating in seven data collection sessions for each participant. In total, we obtained data from 243 sessions, with one participant withdrawing from the last two sessions due to health problems.
\begin{table}[h!]
\centering
\scriptsize 
\caption{Demographics in session 1. NA: Not Available.}
\begin{tabular}{c|ccc|ccc}
\hline
\textbf{Age} & \multicolumn{3}{c|}{\textbf{HC}} & \multicolumn{3}{|c}{\textbf{MCI}} \\
\hline
 & Female & Male & MoCA & Female & Male & MoCA \\
\hline
[65, 70) & 3 & 2 & 27.40 & 2 & 2 & 22.75 \\
\hline
[70, 75) & 5 & 5 & 27.20 & 4 & 5 & 23.89 \\
\hline
[75, 80) & 0 & 0 & NA & 1 & 3 & 23.75 \\
\hline
$\geq$80 & 0 & 0 & NA & 1 & 2 & 21.33 \\
\hline
\textbf{Total} & 8 & 7 & 27.30 & 8 & 12 & 22.93 \\
\hline
\end{tabular}
\label{table:demo}
\end{table}

Participants were asked to answer 18 cognitive-task questions in each session to assess their cognitive abilities. These questions cover various cognitive domains, including naming, memory, attention, sentence repetition, verbal fluency, abstraction, delayed recall, and orientation. The questions were selected from the telephone-based cognitive assessment interviews~\cite{brandt1988telephone}. Examples of these questions are in Table~\ref{tab:evaluation}.
\begin{table}[h]

\centering

\setlength{\tabcolsep}{1pt} 
\scriptsize 
\caption{Cognitive-task questions examples.}

\begin{tabular}{l|l}

\toprule

\hspace{5mm}\textbf{Evaluation} & \hspace{15mm}\textbf{Example Question} \\ \midrule

Naming                        & \textit{"Name the prickly green plant in deserts."} \\ \midrule

Memory                        & \textit{"List as many animals as possible in one minute."} \\ \midrule

Attention                     & \textit{"Count backwards from 20 to 1."} \\ \midrule

Sentence Repetition           & \textit{"Say this, Methodist Episcopal"} \\ \midrule

Verbal Fluency                & \textit{"Name words beginning with 'F' in one minute."} \\ \midrule

Abstraction                   & \textit{"Subtract 7 from 100, then continue subtracting 7."} \\ \midrule

Delayed Recall                & \textit{"Recall the list of 10 words I read earlier."} \\ \midrule

Orientation                   & \textit{"What is the opposite of west?"} \\

                              & \textit{"What is today's date?"} \\ \bottomrule

\end{tabular}

\label{tab:evaluation}

\end{table}

To ensure high-quality data, each participant was provided with a VAS device and given a training session. A research assistant helped each participant use the VAS and record speech. We also advised participants to minimize background noise during their sessions. The VAS was programmed to ask the 18 cognitive-task questions. Participants could use commands such as ``Alexa, next question" and ``Alexa, repeat the question" to control the pace of their answers, allowing for a self-administered process. Participants' cognitive diagnoses were assessed using the MoCA scores, with a cutoff set at 26: scores $\geq 26$ were HC, and scores $<26$ were MCI. On average, each response recording lasted about 30 seconds and was resampled to a 16kHz frequency.

\section{Method}

In this section, we outline the proposed methods, feature extraction, and models. Figure~\ref{fig:pipe} shows the workflow of the two proposed methods, including data collection, feature extraction, fine-tuning, classification models, and outputs. We denote each participant as \( u_i (1\leq i\leq 35) \). The data of a participant \(u_i\) in a virtual session \(s_j (1\leq j\leq 7)\) is \(a_{i,j}\). The time between sessions \(s_j\) and \(s_{j+1}\) is about three months. We propose two methods to detect cognitive state and predict the change at two time points.

\subsection{Cognitive State Detection Method}
\label{sec:problem}
In analyzing the proposed cognitive state detection method, we compare two different approaches.

\subsubsection{Baseline Method}
The baseline method classifies each participant, denoted as $u_i$, as having MCI or being an HC based on data from a single session. Each session is treated independently, without considering the time information of sessions. Results are then averaged across all sessions.

\subsubsection{Proposed Method}
We propose using historical data from all preceding sessions up to the current session as the input. By aggregating the data, represented as \(\{(a_{i,x}) | 1\leq x \leq j\}\), we aim to detect the cognitive state $c_{i,x}$ of \(u_i\) in the session $s_j$ by classifying $c_{i,x}$ as either MCI or HC. We describe this method \(\mathcal{M}_1\) to solve a binary classification problem, where \(\mathcal{M}_1(\{a_{i,x}|1\leq x\leq j\}) = c_{i,x}\). 

\subsection{Cognitive Change Prediction Method}
This proposed method aims to increase the number of data samples for predicting cognitive changes between two different time points, covering short-term and long-term changes. For each participant $u_i$, we select two sessions: $s_{x}$ and $s_y$, where $1\leq x, y\leq 7$. Using the data from these sessions, $a_{i,x}$ and $a_{i,y}$, we predict cognitive changes as a three-class classification problem. The class labels $c_{i,x,y}$ are defined as follows: \textit{improved} if $c_{i,x} = \textbf{MCI}$ and $c_{i,y} = \textbf{HC}$; \textit{decline} if $c_{i,x} = \textbf{HC}$ and $c_{i,y} = \textbf{MCI}$; \textit{no change} if $c_{i,x} = c_{i,y}$. We describe this method as $\mathcal{M}_2$, where $\mathcal{M}_2(a_{i,x}, a_{i,y}) = c_{i,x,y}$. 






\subsection{Feature Extraction}
\subsubsection{Acoustic Features}
\label{sec:acousticfeature}
We extract paralinguistic acoustic features of \textbf{eGeMAPS}~\cite{eyben2015geneva} and \textbf{ComParE-2016}~\cite{eyben2013recent} via the OpenSMILE toolkit~\cite{eyben2010opensmile}. eGeMAPS includes 18 low-level descriptors related to frequency, energy, harmonics-to-noise ratio, and spectral parameters, leading to an 88-dimensional feature vector. ComParE-2016 is a standard feature set from the Interspeech ComParE challenges, leading to a 6,373-dimensional feature vector. Additionally, we apply self-supervised learning models to capture a wide range of acoustic features~\cite{superb, mlsuperb}, including \textbf{HuBERT}~\cite{mottron2006enhanced}, \textbf{WavLM}~\cite{chen2022wavlm}, \textbf{Wav2vec 2.0}~\cite{baevski2020wav2vec}, and \textbf{XLS-R}~\cite{babu2021xls}. Each model contributes uniquely to enhancing speech recognition. HuBERT uses offline clustering to generate pseudo-labels for hidden units, learning both acoustic and language models from inputs. WavLM jointly learns masked speech prediction and denoising, thus using audio sequences to improve recognition. Wav2vec 2.0 jointly learns contextualized speech representations for robust input recognition. XLS-R, a cross-lingual model, excels in large and unlabeled datasets.

\subsubsection{Linguistic Features } \label{sec:linguisticfeature} We use the \textbf{Whisper large-v2}\footnote{\url{https://huggingface.co/openai/whisper-large-v2}} automatic speech recognition (ASR) to transcribe audio recordings to transcripts~\cite{radford2023robust}. Linguistic features are extracted using \textbf{BERT}~\cite{devlin2018bert} and \textbf{LLaMA-2}~\cite{touvron2023llama}. 
Specifically, BERT enables the extraction of token-level embeddings, and we use its last hidden layer to obtain embeddings. We use the uncased base version\footnote{\url{https://huggingface.co/google-bert/bert-base-uncased}}, which consists of 12 transformer layers, 768 hidden units, and around 110 million parameters. We choose the LLaMA-2 7B model\footnote{\url{https://huggingface.co/meta-llama/Llama-2-7b}} to generate embeddings for the transcripts and use the last hidden state for feature extraction. Furthermore, we explore the effects of fine-tuning BERT and LLaMA-2 on our transcripts. After adding a classification layer on top of BERT or LLaMA-2, the training process adjusts the new layer and the last layer of the base model. 

\subsubsection{Fusion of Modality} 
We use a multimodal fusion approach to exploit the strengths of both acoustic and linguistic modalities. We select the top-performing features from each modality and concatenate them into a vector as the input for the model. The fusion modality is expected to improve the classification performance.

\subsection{Models}
\label{sec:models} 
We evaluate four classification models following methods previously used in dementia detection: Decision Tree (DT), Support Vector Machine (SVM), Random Forest (RF), and Neural Networks (NN). For each session, we aggregate each participant's 18 response embeddings and then average them. This approach ensures that the input vector for the models has the same length and remains independent of speakers. 

In the cognitive state detection method, participants completed seven sessions except for one participant who withdrew from two sessions, generating $7\times 34 + 5 = 243$ samples. For the cognitive change prediction method, we randomly pair samples from two different sessions. This approach generates 42 samples for each participant attending 7 sessions, yielding $42\times 34 + 20 = 1448$ samples.

\begin{table*}[htb!]
\centering
\caption{Comparisons between baseline method (see Section~\ref{sec:problem}) and the cognitive state detection method with various types of features. Bold indicates the best results after averaging acoustic features (see Section~\ref{sec:acousticfeature}). Bold+underscore indicates the best results after averaging linguistic features (see Section~\ref{sec:linguisticfeature}).}
\label{tab:1}
\begin{tabular}{lc|cc|cc|cc|cc}
\toprule
\textbf{Feature} & \textbf{Model} & \multicolumn{2}{c|}{\textbf{Acc. (\%)}} & \multicolumn{2}{c|}{\textbf{Prec. (\%)}}& \multicolumn{2}{c|}{\textbf{Recall (\%)}} & \multicolumn{2}{c}{\textbf{F1 (\%)}} \\
 & & \textbf{Baseline}  & \textbf{Proposed}  & \textbf{Baseline} & \textbf{Proposed}  & \textbf{Baseline}  & \textbf{Proposed}  & \textbf{Baseline}  &\textbf{Proposed} \\ 
\toprule
\multirow{4}{*}{\textit{Acoustic}} 
    & DT   & 61.1&  74.7       & 57.0&  75.2      & 59.2  &  74.7       &57.4& 74.3   \\
    & SVM  & \textbf{63.4}&  70.2       &59.2   & 69.0       & 60.5  & 69.8        &\textbf{60.2}  & 69.1  \\
    & RF   & 61.5   &  \textbf{75.2}       & \textbf{61.3}  &  \textbf{75.7}      & 59.5  &\textbf{75.2}       &57.1 &  \textbf{75.2}   \\
    & NN   & 62.5   &  67.7       & 57.7&  70.5     & \textbf{61.4}&  64.8        &59.5  &  66.0  \\  
    \hline
    & \textit{Average} & \textit{62.1}&  \textit{72.0}     & \textit{58.8}&  \textit{72.6}      & \textit{60.2}&  \textit{71.1}       &\textit{58.6}&  \textit{71.2}   \\
\toprule
\multirow{4}{*}{\textit{Linguistic}}   
    & DT & 61.0     &  73.5       &64.8&  67.5      &63.5   &73.5         &59.0  & 68.0   \\
    & SVM  &67.5&  \textbf{\underline{81.1}}       &63.5   & 80.0       &\textbf{\underline{69.1}}  & 78.0        &64.4  & \textbf{\underline{78.0}}   \\
    & RF   &\textbf{\underline{68.5}}     &  79.5       &\textbf{\underline{66.5}}  & \textbf{\underline{81.0}}      &64.5   & 76.0        &\textbf{\underline{64.6}}  & 77.0   \\
    & NN & 62.8     &  79.5        &64.5&78.5       &63.5   &\textbf{\underline{78.5}}         &60.5  &77.5   \\
     \hline
    & \textit{Average} & \textit{65.0}&  \textit{78.4}       & \textit{64.8}&  \textit{76.8}      & \textit{65.2}  &  \textit{76.5}       &\textit{62.1}  &  \textit{75.1}   \\
\toprule
\end{tabular}
\end{table*}

\begin{table*}[!htbp]
\centering
\caption{Comparisons between baseline and cognitive state detection method with fine-tuning BERT and LLaMA-2 model.}
\label{tab:1_bert}
\begin{tabular}{lc|cc|cc|cc|cc}
\toprule
\textbf{ } & \textbf{ } & \multicolumn{2}{c|}{\textbf{Acc. (\%)}} & \multicolumn{2}{c|}{\textbf{Prec. (\%)}}& \multicolumn{2}{c|}{\textbf{Recall (\%)}} & \multicolumn{2}{c}{\textbf{F1 (\%)}} \\
 & & \textbf{Baseline}  & \textbf{Proposed}  & \textbf{Baseline} & \textbf{Proposed}  & \textbf{Baseline}  & \textbf{Proposed}  & \textbf{Baseline}  &\textbf{Proposed} \\ 
\toprule
    \multicolumn{2}{l|}{Fine-tuning BERT}    &57.4& \textbf{84.4}    &\textbf{48.9}& \textbf{82.1}   &\textbf{55.6}& \textbf{80.0}  &48.5& \textbf{81.2}\\
    \midrule
    \multicolumn{2}{l|}{Fine-tuning LLaMA-2}    & \textbf{63.8}  &71.0  &45.7&75.2  &51.8&71.6 & \textbf{51.6}&69.1 \\
\bottomrule
\end{tabular}
\end{table*}

\section{Experimental Setup}
\label{sec: experiments}
We extract acoustic features from ``base models" of self-supervised learning models (see Section~\ref{sec:acousticfeature}). The linguistic feature extraction is via freezing weights, except for the weights in the last hidden layer. Additionally, fine-tuning BERT adopts the BERT-for-Sequence-Classification\footnote{\url{https://github.com/huggingface/transformers}}~\cite{wolf2020transformers}, a learning rate at 2e-05, a warmup of 500 steps, cross-entropy loss, a dropout rate of 0.3, a linear learning rate scheduler, and training 15 epochs with a batch size of 16. For LLaMA-2 fine-tuning, we use the QLoRA (Efficient Fine-tuning of Quantized Large Language Models)~\cite{dettmers2023qlora} on the 7B model. QLoRA quantizes the LLaMA-2 model to 4 bits and freezing parameters, except for a few trainable low-rank adapter layers.

To evaluate the feature extraction method, we use a 10-fold cross-validation with default values of hyperparameters: DT with minimum sample per split of 2, SVM with a radial basis function kernel with penalty 1.0. RF with decision trees set to 100. We apply the min-max normalization to extracted features. Additionally, the NN (Neural Network) classifier is optimized: the Adam optimizer with a learning rate of 0.1, cross-entropy loss, and training 100 epochs with a batch size of 16. We evaluate model performance through accuracy, precision, recall, and F1 scores. All experiments were performed on a single A100 40GB GPU.

\section{Results}
We present the averaged performance of acoustic features from Section~\ref{sec:acousticfeature} and linguistic features from Section~\ref{sec:linguisticfeature} to compare different features when trained with classifiers.

\subsection{Cognitive State Detection Method Performance}
We compare the proposed method against the baseline method in Table~\ref{tab:1}. The finding reveals that incorporating historical data consistently enhances performance. For the acoustic features with all classifiers, there is an increase in average accuracy from 62.1\% to 72.0\% (by 9.9\%) and F1-score from 58.6\% to 71.2\% (by 12.6\%). Similarly, for the linguistic features with all classifiers, there is an increase in average accuracy from 65.0\% to 78.4\% (by 13.4\%) and F1-score from 62.1\% to 75.1\% (by 13.0\%). These results indicate that incorporating historical data effectively improves MCI detection on our small dataset.

Among the acoustic features, HuBERT outperforms other acoustic features. Specifically, HuBERT with DT achieves an accuracy of 81.0\%, precision of 78.3\%, recall of 81.2\%, and an F1-score of 81.1\%. The second-best performance is the ComParE-2016 paralinguistic feature set with SVM and NN models. On the linguistic side, BERT outperforms LLaMA-2, demonstrating an accuracy of 80.0\% accuracy, 77.4\% precision, 80.0\% recall, and 79.1\% F1-score. These results indicate the robustness and adaptability of HuBERT and BERT within this specific domain. HuBERT and BERT effectively identify distinct acoustic and linguistic patterns among individuals with and without MCI. Both acoustic and linguistic features show comparable ability in capturing information for MCI detection.

Table~\ref{tab:1_bert} presents the results of fine-tuning the transcripts.  We observe that fine-tuning BERT leads to an increase in accuracy from 57.4\% to 84.4\% (by 27.0\%) and F1-score from 48.5\% to 81.2\% (by 32.7\%). Fine-tuning LLaMA-2 increases accuracy from 63.8\% to 71.0\% (by 7.2\%) and F1-score from 51.6\% to 69.1\% by (17.5\%). The process of fine-tuning with our proposed method improves the baseline performance, particularly with the BERT model. Results show that BERT outperforms LLaMA-2 across most metrics, indicating that BERT adapts the domain of MCI detection better. LLaMA-2 seems less effective within our dataset.

Table~\ref{tab:2} shows the result of HuBERT and BERT modality fusion with the highest accuracy of 74.9\% and F1-score of 72.7\%. Modality fusion does not improve performance beyond what is achieved using the most effective single modality. This implies that acoustic or linguistic features alone has comparable effectiveness. Adding another modality may not contribute extra information, considering the limitation of small data size.

\begin{table}[!htbp]
\centering
\caption{The result of HuBERT and BERT modality fusion using the cognitive state detection method.}
\label{tab:2}
\begin{tabular}{lcccc}
\toprule
\textbf{Model} & \textbf{Acc. (\%)} & \textbf{Prec. (\%)} & \textbf{Recall (\%)} & \textbf{F1 (\%)} \\
\toprule
{\textit{DT}} 
       &69.6  &66.9  &68.6  &66.5\\
\midrule
{\textit{SVM}}   
       &\textbf{74.9} &\textbf{70.6} &73.6  &69.3 \\
\midrule
{\textit{RF}}   
        &73.4  &67.1 &72.8  &70.2 \\
\midrule
{\textit{NN}}  
      &74.8  &64.8  &\textbf{75.6}  &\textbf{72.7} \\
\bottomrule
\end{tabular}
\end{table}
 

\subsection{Cognitive Change Prediction Performance}

Table~\ref{tab:3} illustrates that using the proposed cognitive change prediction method, acoustic features achieve the highest accuracy of 76.2\% with RF model and the highest F1-score of 73.7\% with DT model. Linguistic features achieve the highest accuracy of 68.7\% with NN model and the highest F1-score of 65.6\% with NN model. These results indicate that our method increases data sample sizes, leading to the effective prediction of cognitive changes. Moreover, acoustic features outperform linguistic features. Specifically, the acoustic HuBERT demonstrates the best performance with the DT model (78.0\% accuracy, 77.8\% precision, 78.2\% recall, and 76.7\% F1-score). Results indicate the effectiveness of acoustic features in identifying cognitive changes within our dataset.

\begin{table}[!htbp]
\caption{Comparisons of the cognitive change prediction method with various types of features. Bold indicates the best results after averaging acoustic features (Section~\ref{sec:acousticfeature}). Bold+underscore indicates the best results after averaging linguistic features (Section~\ref{sec:linguisticfeature}).}
\label{tab:3}
\begin{tabular}{lccccc}
\toprule
\textbf{Feature} & \textbf{Model} & \textbf{Acc. (\%)} & \textbf{Prec. (\%)} & \textbf{Recall (\%)} & \textbf{F1 (\%)} \\
\toprule
\multirow{4}{*}{\textit{Acoustic}} 
    & DT &75.3      &73.8      &\textbf{72.8}      &\textbf{73.7}   \\
   & SVM  &72.7      &70.7      & 69.3          & 64.2   \\
    & RF &\textbf{76.2}     & \textbf{76.0}     &72.7           & 73.1   \\
     & NN &70.2      &70.2      &68.5          & 67.3  \\
     \hline
    & \textit{Average} & \textit{73.6}      & \textit{72.7}      & \textit{70.8}  &  \textit{69.6}  \\
\toprule
\multirow{4}{*}{\textit{Linguistic}}   
    & DT  & 64.5      & 63.0      &62.5     & 61.5   \\
    & SVM  &68.5      &\textbf{\underline{66.0}}     &\textbf{\underline{68.5}}      &62.2   \\
    & RF &66.5      &62.0      & 66.5      &64.5   \\
     & NN &\textbf{\underline{68.7}}      &64.5     & 67.0      &\textbf{\underline{65.6}}   \\
   \hline
   & \textit{Average} & \textit{67.1}     & \textit{63.9}      & \textit{66.1}          &  \textit{63.5}  \\
\toprule
\end{tabular}
\end{table}

Table~\ref{tab:3_bert} shows that fine-tuning BERT achieves the accuracy of 70.0\% and F1-score of 68.0\%, while LLaMA-2 achieves the accuracy of 59.1\% and F1-score of 56.9\%. These results show that BERT outperforms LLaMA-2 across all metrics, indicating that BERT is effective and adapts better in identifying cognitive changes within our dataset. Moreover, fine-tuning the transcripts outperforms the best feature extraction method with NN model, leading to an increase from 68.7\% to 70.0\% (by 1.3\%) in accuracy and 65.6\% to 68.0\% (by 2.4\%) in F1-score. However, acoustic features still outperform the fine-tuned BERT performance.

\begin{table}[!htbp]
\caption{Results of the cognitive change prediction method with fine-tuning BERT and LLaMA-2 model.}
\label{tab:3_bert}
\begin{tabular}{lccccc}
\toprule
\textbf{ } & \textbf{ } & \textbf{Acc. (\%)} & \textbf{Prec. (\%)} & \textbf{Recall (\%)} & \textbf{F1 (\%)} \\
\toprule
    \multicolumn{2}{l}{Fine-tuning BERT}    &\textbf{70.0}   &\textbf{68.4}  &\textbf{70.7}  &\textbf{68.0}  \\
    \midrule
    \multicolumn{2}{l}{Fine-tuning LLaMA-2}    &59.1  &48.6  &59.4 &56.9  \\
\bottomrule
\end{tabular}
\end{table}

Table~\ref{tab:4} presents the results of HuBERT and BERT modality fusion. This method achieves the highest accuracy of 76.0\% and the highest F1-score of 71.6\%. Compared to the best linguistic feature extraction (NN model), modality fusion improves from 68.7\% to 76.0\% (by 7.3\%) in accuracy and from 65.6\% to 71.6\% (by 6.0\%) in F1-score. This indicates that modality fusion can effectively enhance model performance compared to the single modality in predicting cognitive changes when a larger set of data samples is available.

\begin{table}[!htbp]
\caption{The result of HuBERT and BERT modality fusion using the cognitive change prediction method.}
\label{tab:4}
\centering
\begin{tabular}{lcccc}
\toprule
\textbf{Model} & \textbf{Acc. (\%)} & \textbf{Prec. (\%)} & \textbf{Recall (\%)} & \textbf{F1 (\%)} \\
\toprule
{\textit{DT}} 
      &69.1  &70.0  &67.5  &70.0  \\
\midrule
{\textit{SVM}}   
       &64.5  &62.3  &63.6  &69.3  \\
\midrule
{\textit{RF}}   
         &\textbf{76.0}  &\textbf{73.3}  &71.2  &\textbf{71.6}\\
\midrule
{\textit{NN}}  
        &70.2  &71.8  &\textbf{71.4}  &69.0    \\
\bottomrule
\end{tabular}
\end{table}

\section{Conclusion} 
This paper presents a longitudinal study using VAS to collect cognitive-task speech sessions for early detection of cognitive decline. We evaluate two proposed methods that employ the advantages of longitudinal data: the cognitive state detection method and the cognitive change prediction method. Our findings reveal that methods incorporating historical data consistently outperform those that do not. We also explore the effectiveness of acoustic and linguistic features in capturing distinguishable information related to MCI. The results show that both acoustic and linguistic features have similar capabilities in identifying MCI. Specifically, acoustic features from HuBERT and linguistic features from BERT achieved the best performance in their respective categories. However, in our experiment, the modality fusion of acoustic and linguistic features does not enhance the performance of the acoustic features alone. The limitations of our work include a small participant group, unbalanced data labels, and limited cognitive changes observed during the study period. Future work strives to extend our study design with more participants and longer periods, expand task questions, and include natural conversations to enhance the accurate assessment of cognitive and functional abilities.

 \section{Acknowledgement}
This research is funded by the US National Institutes of Health National Institute on Aging R01AG067416. We thank Professor David Kotz and Professor Brian MacWhinney for providing guidance.


\bibliographystyle{IEEEtran}
\bibliography{mybib}

\end{document}